\documentstyle[preprint,epsfig,aps]{revtex}
\begin{document}

\preprint{\vbox{\hbox{IFT--P.034/97}    \vspace{-.3cm}
                \hbox{IFUSP 1264}       \vspace{-.3cm}
                \hbox{FTUV/97--20}      \vspace{-.3cm}
                \hbox{IFIC/97--20}      \vspace{-.3cm}
                \hbox{hep-ph/9704400}}}

\title{Limits on Anomalous Top Couplings from \boldmath{$Z$} Pole Physics}

\author{O.\ J.\ P.\ \'Eboli$^1$, M.\ C.\ Gonzalez-Garcia$^{2,3}$, and 
        S.\ F.\ Novaes$^2$}

\address{$^1$ Instituto de F\'{\i}sica ,
              Universidade de S\~ao Paulo \\
              C.\ P.\ 66.318, 05315--970 --  S\~ao Paulo, SP, Brazil.}

\address{$^2$ Instituto de F\'{\i}sica Te\'orica, 
              Universidade Estadual Paulista \\   
              Rua Pamplona, 145,
              01405--900 --  S\~ao Paulo, Brazil} 

\address{$^3$ Instituto de F\'{\i}sica Corpuscular --  IFIC/CSIC,
              Departament de F\'{\i}sica Te\`orica \\
              Universitat de Val\`encia, 46100 Burjassot, Val\`encia, Spain}

\date{\today} 

\maketitle

\begin{abstract} 
  We obtain constraints on possible anomalous interactions of the top
  quark with the electroweak vector bosons arising from the precision
  measurements at the $Z$ pole. In the framework of $SU(2)_L \otimes
  U(1)_Y$ chiral Lagrangians, we examine all effective CP--conserving
  operators of dimension five which induce fermionic currents
  involving the top quark. We constrain the magnitudes of these
  anomalous interactions by evaluating their one--loop contributions
  to the $Z$ pole physics. Our analysis shows that the operators that 
  contribute to the LEP observables get bounds close to the theoretical
  expectation for their anomalous couplings. We also show that those which
  break the $SU(2)_C$ custodial symmetry are more strongly bounded.
\end{abstract}


\section{Introduction}

The Standard Model (SM) of electroweak interactions has passed through
an intense experimental scrutiny that confirmed several of its
predictions. In particular, the precise LEPI measurements performed at
the $Z$ pole show that the SM describes extremely well the couplings
between the gauge bosons and the light fermions \cite{ewwg}.
Notwithstanding, the couplings of the top quark to the gauge bosons
are still rather poorly measured at the Tevatron $p\bar p$ collider
\cite{teva}.  Furthermore, some other elements of the SM, such as the
symmetry breaking mechanism, have not been directly tested yet.

If the breaking of the $SU(2)_L \otimes U(1)_Y$ symmetry takes place
via the Higgs mechanism with a relatively light elementary Higgs
boson, both the symmetry breaking and the fermion mass generation can
have a common origin. However, if no fundamental Higgs particle is
present in the theory, the mechanism that breaks the electroweak
symmetry and the one that gives rise to the fermion masses are not
necessarily related, and we can envisage a breaking in the
universality of the fermionic interactions \cite{peccei}. One may
expect that the top quark, being the heaviest of the known fermions,
should be more sensitive to the existence of new physics in the
electroweak breaking sector. This is certainly the case if, for
instance, the breaking of the electroweak symmetry occurs dynamically
via the appearance of a $t \bar{t}$ condensate \cite{tec}.

Whatever the dynamics of the symmetry breaking mechanism is,
renormalizability requires that this breaking must occur
spontaneously.  This leads to the existence of Goldstone bosons
associated with the broken directions which become the longitudinal
components of the massive gauge bosons. Assuming this as our starting
point, we can build effective low--energy Lagrangians which describe
the interactions of these Goldstone bosons. The self--interactions of
the Goldstone bosons, to lowest order, are totally determined by the
symmetry breaking pattern and it is described in terms of a unique
dimensionful parameter $v$.  However, the interactions between the
Goldstone bosons and other fields, such as fermions, involve new
unknown parameters that, when the interaction is gauged, leads, in
general, to universality violation in the couplings between gauge
boson and fermions.

Limits on universality violation in the interactions of the top quark
to the gauge bosons have been studied before in Ref.\
\cite{peccei,yuan} where the authors included only dimension--four
operators. In this work, we study the most general CP invariant
dimension--five Lagrangian for the interactions between the Goldstone
bosons and the top and bottom quarks. In the unitary gauge, these
Lagrangians give rise to non--universal couplings of
the top and bottom quarks to the gauge bosons. Since the SLC and LEPI
achieved a precision of the order of $0.1$ percent in some
observables, the $Z$ pole physics is the best available source of
information on these interactions. We obtain the constraints on these
anomalous top couplings by imposing that their one--loop contributions
to the electroweak parameters are compatible with the $Z$ pole data
\cite{altarelli}.


\section{Effective Lagrangians}
 
If the Higgs boson, responsible for the electroweak symmetry breaking,
is very heavy, it can be effectively removed from the physical
low--energy spectrum. In this case and for dynamical symmetry breaking
scenarios relying on new strong interactions, one is led to consider
the most general effective Lagrangian which employs a nonlinear
representation of the spontaneously broken $SU(2)_L \otimes U(1)_Y$
gauge symmetry \cite{appelquist}. The resulting chiral Lagrangian is a
non--renormalizable nonlinear $\sigma$--model coupled in a
gauge--invariant way to the Yang-Mills theory.  This model independent
approach incorporates by construction the low--energy theorems
\cite{cgg}, that predict the general behavior of Goldstone boson
amplitudes, irrespective of the details of the symmetry breaking
mechanism. Unitarity requires that this low--energy effective theory
should be valid up to some energy scale smaller than $4\pi v \simeq 3$
TeV, where new physics would come into play.

In order to specify the effective Lagrangian for the Goldstone bosons,
we assume that the symmetry breaking pattern is $G = SU(2)_L \otimes
U(1)_Y$ $\longrightarrow$ $H = U(1)_{\text{em}}$, leading to just
three Goldstone bosons $\pi^a$ ($a=1,2,3$). With this choice, the
building block of the chiral Lagrangian is the dimensionless
unimodular matrix field $\Sigma$,
\begin{equation}
\Sigma = \exp \left(i\frac{\pi^a\tau^a}{v}\right) \; ,
\end{equation}
where $\tau^a$ ($a=1,2,3$) are the Pauli matrices.  We implement the
$SU(2)_C$ custodial symmetry by imposing a unique dimensionful
parameter, $v$, for charged and neutral fields. Under the action of
$G$ the transformation of $\Sigma$ is
\begin{displaymath}
\Sigma \rightarrow \Sigma' = L~ \Sigma~ R^\dagger \; ,
\end{displaymath}
with
$L= \exp \left(i\frac{\alpha^a\tau^a}{2}\right)$ and $R =
\exp\left(iy \frac{\tau^3}{2}\right)\;$.  
$\alpha^{a}$ and $y$ are the parameters of the transformation.

The gauge fields are represented by the matrices $\hat{W}_{\mu} =
\tau^a W^a_{\mu}/(2i)$, $\hat{B}_{\mu} = \tau^3 B_{\mu}/(2i)$, while
the associated field strengths are given by
\begin{equation} 
\begin{array}{ll}
\hat{W}_{\mu\nu} =& \partial_{\mu}\hat{W}_{\nu} - 
\partial_{\nu}\hat{W}_{\mu} -g\left[\hat{W}_{\mu},\hat{W}_{\nu}\right]\; ,
\\
\hat{B}_{\mu\nu} =& \partial_{\mu}\hat{B}_{\nu} - \partial_{\nu}
\hat{B}_{\mu} \; .
\end{array}
\end{equation} 
In the nonlinear representation of the gauge group $SU(2)_L \otimes
U(1)_Y$, the mass term for the vector bosons is given by the lowest
order operator involving the matrix $\Sigma$. Therefore, the kinetic
Lagrangian for the gauge bosons reads
\begin{eqnarray}
{\cal L}_B = \frac{1}{2} {\rm Tr} \left( \hat{W}_{\mu\nu}
\hat{W}^{\mu\nu} + \hat{B}_{\mu\nu} \hat{B}^{\mu\nu}\right) +
\frac{v^2}{4}{\rm Tr}\left(D_{\mu}
\Sigma^{\dagger}D^{\mu}\Sigma\right)\;,
\label{mass}
\end{eqnarray}
where the covariant derivative of the field $\Sigma$ is 
$ D_{\mu}\Sigma = \partial_{\mu}\Sigma - g\hat{W}_{\mu}\Sigma 
+ g' \Sigma \hat{B}_{\mu}$.

In order to include fermions in this framework, we must define
their transformation under $G$.  Following Ref.\ \cite{peccei}, we
postulate that matter fields feel directly only the electromagnetic
interaction $f \rightarrow f' = e^{iyQ_f}\;f $,
where $Q_f$ stands for the electric charge of fermion $f$.
The usual fermion doublets are then defined with the following transformation
under $G$  
\begin{equation}
\Psi_L = \Sigma \left(\begin{array}{c}f_1\\[-0.1cm] f_2\end{array}\right)_L 
\; \rightarrow \Psi^\prime_L = L~ \exp(i y Y /2) \Psi_L ,
\end{equation}
where $Q_{f_1} - Q_{f_2} = 1$ and
with $Y = 2 Q_{f_1} -1$.  Right--handed fermions are just the singlets
$f_R$. In this framework, the lowest--order interactions between
fermions and vector bosons that can be built are of dimension four,
leading to anomalous vector and axial--vector couplings, which were
analyzed in detail in Ref.\ \cite{yuan}.

In order to construct the most general Lagrangian describing these
interactions, it is convenient to define the vector and tensor fields
\begin{equation}
\begin{array}{l}
\Sigma_{\mu}^a = -\frac{\displaystyle i}{\displaystyle 2}
{\rm Tr}\left(\tau^a V_\mu^R\right) 
~=~ -\frac{\displaystyle i}{\displaystyle 2}{\rm Tr}
\left(\tau^a\Sigma^{\dagger} D_{\mu}\Sigma\right) \; ,
\\
\Sigma_{\mu\nu}^a = -i\;{\rm Tr}\left[\tau^a \Sigma^{\dagger}
\left[D_{\mu},D_{\nu}\right]\Sigma\right] \; .
\end{array}
\end{equation}
Under $G$, $\Sigma_{\mu}^3$ and $\Sigma_{\mu\nu}^3$ are
invariant while $ \Sigma_{\mu(\mu\nu)}^{\pm} \rightarrow
{\Sigma'}_{\mu(\mu\nu)}^{\pm} = e^{\pm iy}
\Sigma_{\mu(\mu\nu)}^{\pm}$, where $\Sigma_{\mu(\mu\nu)}^{\pm} =
(1/\sqrt{2}) (\Sigma_{\mu(\mu\nu)}^1 \mp i\Sigma_{\mu(\mu\nu)}^2)$.

The basic fermionic elements for the construction of neutral- 
and charged-current effective interactions are
\begin{equation}
\begin{array}{ll}
 \Delta_X (q, q^\prime) \equiv \bar{q} P_X q^\prime 
\; , \;\;
&\Delta^{\mu}_X (q, q^\prime) \equiv \bar{q}  P_X \tilde D^\mu
q^\prime\; , \\
 \overline{\Delta^{\mu}_X}(q,q^\prime)
\equiv\overline{\tilde D^\mu q}P_X  q^\prime\; , \;\;\
&\Delta^{\mu\nu}_X (q, q^\prime) \equiv \bar{q} \sigma^{\mu\nu} P_X
q^\prime   \; ,
\label{fer}
\end{array}
\end{equation}
where $P_X$ ($X=0$, $5$, $L$, and $R$) stands for $I$, $\gamma^5$,
$P_L$, and $P_R$ respectively, with $I$ being the identity matrix and
$P_{L(R)}$ the left (right) chiral projector. The fermionic field $q$
($q^\prime$) represents any quark flavor. $\tilde D^\mu$ represents the
electromagnetic covariant derivative.

The most general neutral--current interactions of dimension--five,
which are invariant under nonlinear transformations under $G$, are
\cite{yuan2}:
\begin{equation}
\begin{array}{ll}
{\cal L}^{\text{NC}} = & a_1^{\text{NC}} ~\Delta_0(t,t)~  
\Sigma_\mu^+  \Sigma^{- \mu} 
+ a_2^{\text{NC}} ~ \Delta_0(t,t)~  \Sigma_\mu^3 \Sigma^{3 \mu}
+i~ a_3^{\text{NC}} ~ \Delta_5(t,t)~\partial^\mu \Sigma^{3}_\mu \\ 
& + i~ b_1^{\text{NC}} ~ \Delta^{\mu\nu}_0 (t,t) ~
\mbox{Tr} \left [ T \hat{W}_{\mu\nu} \right ] 
+ b_2^{\text{NC}} ~ \Delta^{\mu\nu}_0 (t,t)~ B_{\mu\nu} \\
& + i~ b_3^{\text{NC}}~ \Delta^{\mu\nu}_0 (t,t)~ 
\left ( \Sigma^+_\mu \Sigma^-_\nu - \Sigma^+_\nu \Sigma^-_\mu
\right )~
 + i~ c_{1}^{\text{NC}}~\left(\Delta^{\mu}_0 (t,t)-\overline{\Delta^{\mu}_0}
(t,t)\right)\Sigma^{3 \mu} \; ,
\end{array}
\label{nc}
\end{equation}
and the charged--current interactions are
\begin{equation}
\begin{array}{ll}
{\cal L}^{\text CC} = & a_{1L}^{\text{CC}}~ \Delta_L(t,b)~  
\Sigma_\mu^+ \Sigma^{3 \mu} 
+ a_{1R}^{\text{CC}} ~ \Delta_R(t,t)~  \Sigma_\mu^+ \Sigma^{3 \mu} \\
 & +i a_{2L}^{\text{CC}}~ \Delta_L(t,b)~  
\tilde D^\mu  \Sigma_\mu^+ 
+i a_{2R}^{\text{CC}} ~ \Delta_R(t,t)~  \tilde D^\mu \Sigma_\mu^+ \\  
& + \, b_{1L}^{\text{CC}} \Delta^{\mu\nu}_L (t,b) ~
\Sigma_{\mu\nu}^+ 
 +\, b_{1R}^{\text{CC}}~ \Delta^{\mu\nu}_R (t,b) ~
\Sigma_{\mu\nu}^+ \\
& +\, b_{2L}^{\text{CC}}~ \Delta^{\mu\nu}_L (t,b)~ 
\left ( \Sigma^+_\mu \Sigma^3_\nu - \Sigma^+_\nu \Sigma^3_\mu
\right ) 
+\, b_{2R}^{\text{CC}}~ \Delta^{\mu\nu}_R (t,b)~ 
\left ( \Sigma^+_\mu \Sigma^3_\nu - \Sigma^+_\nu \Sigma^3_\mu
\right ) \\
& + i~ c_{1L}^{\text{CC}}~\Delta^{\mu}_L (t,b)\Sigma^{+}_\mu 
+ i~ c_{1R}^{\text{CC}}~\Delta^{\mu}_R (t,b)\Sigma^{+}_\mu 
~+~ \mbox{h.c.} \; .
\end{array}
\label{cc}
\end{equation}

In the unitary gauge, we can rewrite these interactions as a scalar
(${\cal L}_S$), a vector (${\cal L}_V$), and a tensorial (${\cal
  L}_T$) Lagrangian involving the physical fields.
\begin{equation} 
\begin{array}{ll}
{\cal L}_S= & \frac{\displaystyle g^2}{\displaystyle 4 \Lambda}
\Biggl[    \bar{t} t \Bigl(
2\alpha_1^{NC} W^+_\mu W^{-\mu} + \frac{\alpha_2^{NC}}{c_W^2} Z^\mu Z_\mu
\Bigr) \Biggr] + i \frac{\displaystyle g}{\displaystyle 2 c_W \Lambda}
\alpha_3^{NC} \bar{t} \gamma^5 t\partial^\mu Z_\mu\\
& +\frac{\displaystyle g^2}{\displaystyle 2 \sqrt{2} \Lambda c_W} \Biggl\{
\bar{t}~ \Bigl [\alpha_{1L}^{CC} (1-\gamma^5) + \alpha_{1R}^{CC}
(1+\gamma^5) \Bigr]b\; 
 W_\mu^+ Z^\mu \\
 & +\bar{b} \Bigl [\alpha_{1L}^{CC} (1+\gamma^5) +
 \alpha_{1R}^{CC}(1-\gamma^5)\Bigr]t \; 
 W^-_\mu Z^\mu \Biggr\} \\
 & +i \frac{\displaystyle g}{\displaystyle 2 \sqrt{2} \Lambda } \Biggl\{
 \bar{t}~ \Bigl [\alpha_{2L}^{CC} (1-\gamma^5) + \alpha_{2R}^{CC}
(1+\gamma^5) \Bigr]b \;
\left(\partial^\mu W_\mu^+ + i e A^\mu W_\mu^+\right) \\ 
& -\bar{b}~ \Bigl [\alpha_{2L}^{CC} (1+\gamma^5) + \alpha_{2R}^{CC}
(1-\gamma^5) \Bigr]t \;
\left(\partial^\mu W_\mu^- - i e A^\mu W_\mu^-\right)\Biggr\} 
\\
\end{array}
\label{lags}
\end{equation}
\begin{equation}
\begin{array}{ll}
{\cal L}_V= & i ~\frac{\displaystyle  g}{\displaystyle  2 c_W}~ 
\gamma^{NC} ~\bar{t}  ~(\tilde D_\mu t) ~ Z^\mu 
 - i ~\frac{\displaystyle  g}{\displaystyle  2 c_W}~ 
\gamma^{NC} \overline {(\tilde D_\mu t)} 
 ~t ~Z^\mu\\
& +i ~\frac{\displaystyle  g}{\displaystyle  2 \sqrt{2}}~ 
\bar{t} \left [ \gamma_L^{CC} (1-\gamma^5) + \gamma_R^{CC} (1+\gamma^5)\right]~
(\tilde D_\mu b) ~ W{+^\mu} \\
& - i ~\frac{\displaystyle  g}{\displaystyle  4 c_W}~ 
\overline{(\tilde D_\mu b)} 
\left [ \gamma_L^{CC} (1+\gamma^5) + \gamma_R^{CC} (1-\gamma^5)\right]~t ~
 W^{-\mu}
\end{array}
\label{lagv}
\end{equation}
\begin{equation}
\begin{array}{ll}
{\cal L}_T= & \frac{\displaystyle 1}{\displaystyle 4 \Lambda}
\Biggl[ \bar{t} \sigma^{\mu\nu}t \;\Bigl 
( \beta_1^{NC} e F_{\mu\nu} + \beta_2^{NC} \frac{g}{c_W} Z_{\mu\nu}
+4 i g^2  \beta_3^{NC} W_\mu^+ W_\nu^{-} \Bigr) \Biggr] 
\\
& +\frac{\displaystyle g}{\displaystyle 2 \sqrt{2} \Lambda} 
\Biggl\{
\bar{t} \;\sigma^{\mu\nu}\;\Bigl [\beta_{L1}^{CC}(1-\gamma^5)+\beta_{R1}^{CC}
(1+\gamma^5) \Bigr] b \;
\Bigl[W^+_{\mu\nu}+i e\left(A_\mu W_\nu^+-A_\nu W_\mu^+\right) \Bigr]
\\
& +
\bar{b} \;\sigma^{\mu\nu} \;\Bigl [\beta_{L1}^{CC}(1+\gamma^5)+\beta_{R1}^{CC}
(1-\gamma^5) \Bigr] t \;
\Bigl[W^-_{\mu\nu}-i e\left(A_\mu W^-_\nu-A_\nu W^-_\mu\right) \Bigr]
\\
&
+i \frac{\displaystyle g}{\displaystyle c_W}~
\bar{t} \;\sigma^{\mu\nu} \;\Bigl [\beta_{L2}^{CC}(1-\gamma^5)+\beta_{R2}^{CC}
(1+\gamma^5) \Bigr]b\;
\left(Z_\mu W_\nu^+-Z_\nu W_\mu^+\right)
\\
& -i\frac{\displaystyle g}{\displaystyle c_W}~
\bar{b}\; \sigma^{\mu\nu} \;\Bigl
[\beta_{L2}^{CC}(1+\gamma^5)+\beta_{R2}^{CC}
(1-\gamma^5) \Bigr] t  \;
\left(Z_\mu W^-_\nu-Z_\nu W^-_\mu\right)\; 
 \Biggr\}
 \\
\end{array}
\label{lagt}
\end{equation}
The couplings constants $\alpha$'s, $\beta$'s and $\gamma$'s are
linear combinations of the $a$'s, $b$'s and $c$'s in Eqs.\ (\ref{nc})
to (\ref{cc}).  In writing the interactions (\ref{lags}) and
(\ref{lagt}), the coupling constants were defined in such a way that
we have a factor $g/(2c_W)$ per $Z$ boson, $g/\sqrt{2}$ per $W^\pm$,
and $e$ per photon. $s_W$ ($c_W$) is the sine (cosine) of the weak mixing 
angle, $\theta_W$.
Similar interactions were obtained in Ref.\ 
\cite{yuan2}\footnote{ Notice that we agree with Ref.\ \cite{yuan2} in
  the number of NC interactions (7) but we have only 10 CC
  interactions since the Lagrangian in Eq.\ (62) of this reference can
  be reduced to Eq.\ (61) and Eq.\ (64) up to a total derivative.},
  and for a linearly realized symmetry group, in Ref.\ \cite{gouna}.

In general, since chiral Lagrangians are related to strongly
interacting theories, it is hard to make firm statements about the
expected order of magnitude of the couplings. Notwithstanding,
requiring the loop corrections to the effective operators to be of the
same order of the operators themselves suggests that these
coefficients are of ${\cal O}(1)$ \cite{wudka}. Moreover, if the high
energy theory respects chiral symmetry, we can also foresee a further
suppression factor proportional to $m_{\text{top}}/\Lambda$.

As an example of the above anomalous couplings, we show their
couplings for the SM with a heavy Higgs boson integrated out. In this
case, we can perform the matching between the full theory and the
effective Lagrangian \cite{dittmaier}. Setting $m_b=0$ and keeping only the
leading terms of the order $m_{\text{top}} \log(M_H^2)$, we find that
only two effective operators are generated
\begin{equation}
\alpha_1^{NC}=\alpha_2^{NC}=
\frac{g^2 m_{\text{top}}\Lambda}{16\pi^2 M_W^2}
\, \log\frac{M_H^2}{m^2_{\text{top}}}\; .
\end{equation}


\section{Limits from $Z$ Pole Physics}

At the one-loop level, the effective interactions (\ref{lags}) to
(\ref{lagt}) contribute to the $Z$ physics through universal
corrections to the gauge boson propagators and non--universal ones to
the $Z b\bar{b}$ vertex.  The oblique anomalous corrections can be
efficiently summarized in terms of the parameters 
$\epsilon^1_{\text{new}}$, $\epsilon^2_{\text{new}}$, and
$\epsilon^3_{\text{new}}$ \cite{abc}, whose expressions as functions
of the unrenormalized gauge boson self-energies in the on--mass--shell
renormalization scheme are
\begin{eqnarray*}
\epsilon^1_{\text{new}} &=& \frac{\Sigma_{\text{new}}^{ZZ}(M_Z^2)}{M_Z^2}
- \frac{\Sigma_{\text{new}}^{WW}(0)}{M_W^2}
+ 2 \; \frac{s_W}{c_W} \; \frac{\Sigma_{\text{new}}^{AZ}(0)}{M_Z^2}
-\Sigma_{\text{new}}^{\; \prime \; ZZ}(M_Z^2)\; ,
\nonumber \\
\epsilon^2_{\text{new}} &=& \frac{\Sigma_{\text{new}}^{WW}(M_W^2) - 
\Sigma_{\text{new}}^{WW}(0)}{M_W^2}
- s_W^2 \; \frac{\Sigma_{\text{new}}^{AA}(M_Z^2)}{M_Z^2} \nonumber \\
&& \; - 2 s_W c_W \left [ \frac{\Sigma_{\text{new}}^{AZ}(M_Z^2)  - 
\Sigma_{\text{new}}^{AZ}(0)}{M_Z^2} \right ] - 
c_W^2 \; \Sigma_{\text{new}}^{\; \prime \; ZZ}(M_Z^2) \; ,
\nonumber \\
\epsilon^3_{\text{new}} &=& c_W^2 \; 
\frac{\Sigma_{\text{new}}^{AA}(M_Z^2)}{M_Z^2} + (c_W^2 - s_W^2)\;
\frac{c_W}{s_W} \; \left[ \frac{\Sigma_{\text{new}}^{AZ}(M_Z^2)  - 
\Sigma_{\text{new}}^{AZ}(0)}{M_Z^2} \right] - 
c_W^2 \; \Sigma_{\text{new}}^{\; \prime \; ZZ}(M_Z^2) \; ,
\end{eqnarray*}
where $\Sigma_{\text{new}}^{V_1V_2}$ is the new physics contribution
to the transverse part of $V_1-V_2$ vacuum polarization, and
$\Sigma_{\text{new}}^{\; \prime} \equiv d\Sigma_{\text{new}}/dq^2$.
The above expressions are valid for an arbitrary momentum dependence
of the vacuum polarization diagrams.

We parametrize the anomalous non--universal contributions to the vertex
$Z b \bar{b}$ as
\begin{equation}
i \frac{e}{2 s_W c_W} \left ( \gamma_\mu F_{V}^{Zb} - 
\gamma_\mu \gamma_5 F_{A}^{Zb}  \right ) \; .
\end{equation}
Our results show that the new operators lead to pure left-handed
contributions to this vertex, {\em i.e.} $F_{V}^{Zb}=F_{A}^{Zb}$, in
the limit of vanishing bottom quark mass. These corrections can be
cast in terms of the $\epsilon_b$ parameter \cite{abc,berna}
\begin{equation}
\epsilon^b_{\text{new}} = - 2~ F_{V}^{Zb} \; .
\end{equation}

Recent global analyses of the LEP, SLD, and low-energy data yield the
following values for the oblique parameters \cite{altarelli}, which
include the standard model and new physics contributions, {\it i.e.\/}
$\epsilon^i \equiv \epsilon^i_{\text SM} + \epsilon^i_{\text{new}}$
($i = 1,2,3,b$)
\begin{equation}
\begin{array}{ll}
\epsilon^1 ~=~ (4.28\pm 1.25)\times 10^{-3} \;\; , & \;\;\;\;\;\;\;\;  
\epsilon^2 ~=~ (-7.85\pm 2.2)\times 10^{-3}
\;\; , \\
\epsilon^3 ~=~ (4.13\pm 1.37)\times 10^{-3} \;\; , & \;\;\;\;\;\;\;\;  
\epsilon^b ~=~ (-4.45\pm 3.)\times 10^{-3}
\; .
\end{array}
\label{epsilons}
\end{equation}

In order to include low-energy observables in the extraction of the
values for the $\epsilon$'s, one must assume that the vacuum
polarization corrections differ from the SM ones only by terms up to
order $q^2$ in the momentum expansion.  Since this is the case for the
couplings we are considering, we are allowed to use the values in Eq.\
(\ref{epsilons}) in our analysis. The extraction of the values of the
$\epsilon$ parameters due to new physics requires the subtraction the
SM contribution, which depends upon the SM parameters, and in
particular, on the top quark mass $m_{\text{top}}$.

Our procedure to obtain the bounds on the operators (\ref{lags}) to
(\ref{lagt}) is the following: first we evaluate their corrections to
the gauge boson self--energies and to the $Z b \bar{b}$ vertex using
dimensional regularization \cite{dim}, and neglecting the external
fermion masses. Then, we use the leading non--analytic contributions
from the loop diagrams to constrain the new interactions --- that is,
we keep only the logarithmic terms, dropping all the others. The
contributions that are relevant for our analysis are easily obtained
by the substitution
\[
\frac{2}{4-d} \rightarrow {\rm{log}}\;\frac{\Lambda^2}{\mu^2}\; ,
\]
where $\Lambda$ is the energy scale which characterizes the appearance
of new physics, and $\mu$ is the scale in the process, which we take
to be $\mu=m_{\text{top}}$.

The contributions to the oblique parameters due to the top anomalous
interactions are
\begin{equation}
\begin{array}{ll}
\epsilon^1_{\text{new}} =& 
\frac{\displaystyle g^2}{\displaystyle 96  \pi^2}~
\frac{\displaystyle m^3_{\text{top}} }
{\displaystyle \Lambda~ M_W^2} ~N_c~
\left[12\left(\alpha_2^{NC} - \alpha_1^{NC} \right)
+(12-32 s_W^2)\gamma^{NC} -3\gamma_L^{CC}
\right]
\log\frac{\displaystyle\Lambda^2}{\displaystyle \mu^2} 
\; , \\
\epsilon^2_{\text{new}}= & 
\frac{\displaystyle g^2}{\displaystyle 96  \pi^2}~
 \frac{\displaystyle m_{\text{top}} }{\displaystyle \Lambda} ~ N_c
 \left[
6\left(2\beta_{L1}^{CC} - \beta_2^{NC} - \beta_1^{NC} s_W^2\right)
+2 \gamma_L^{NC} -\gamma_L^{CC}
 \right]\log\frac{\displaystyle\Lambda^2}{\displaystyle \mu^2} 
\; , \\
\epsilon^3_{\text{new}}= & 
\frac{\displaystyle g^2}{\displaystyle 288  \pi^2}~
\frac{\displaystyle m_{\text{top}}}{\displaystyle \Lambda}~ N_c~ 
\left[
3\left(3~ \beta_1^{NC} + 2~ \beta_2^{NC} + 2~\beta_1^{NC} s_W^2\right)~
-2 \gamma^{NC}
\right] \log\frac{\displaystyle\Lambda^2}{\displaystyle \mu^2} \; ,
\end{array}
\label{e123}
\end{equation}
where $N_c = 3$ is the number of colors. 

The anomalous contributions to the $Z b\bar{b}$ vertex are left-handed
for $m_b = 0$, and their expression in terms of the $\epsilon_b$
parameter is
\begin{equation}
\begin{array}{ll}
\epsilon^b_{\text{new}}= &
\frac{\displaystyle g^2}{ \displaystyle 32  \pi^2}~
\frac{\displaystyle m_{\text{top}}^3}{\displaystyle \Lambda M_W^2} 
\Biggl\{
\left(1-3 X_W\right)
\left(\alpha_{1L}^{CC} + 6~ \beta_{L2}^{CC} - 6~ \beta_{L1}^{CC} c_W^2\right) 
 \\
  &+  ~ \alpha_{2L}^{CC} 
\left[1+c_W^2-\frac{\displaystyle s_W^2} {\displaystyle 9 c_W^2 }X_W
(8-27 c_W^2)\right] 
~ + ~ \left(2+ 3 X_W\right)
\gamma^{NC} \\ 
 & ~-\frac{ \displaystyle 1 }{\displaystyle 3}~
 \gamma_{L}^{CC} 
\left[1+2 c_W^2-\frac{\displaystyle 1}{\displaystyle 3 c_W^2}
X_W(11-5c_W^2-18c_W^4)\right] \Biggr\}
\log\frac{\displaystyle\Lambda^2}{\displaystyle\mu^2} \\   \; .
\label{eb}
\end{array}
\end{equation}
where $X_W=M_W^2/m_{\text{top}}^2$.  We made a consistency check of
our calculation by analyzing the effect of these new interactions to
the $\gamma b \bar{b}$ vertex at zero momentum, which is one of the
renormalization conditions in the on--shell renormalization scheme. We
verified that our result for this vertex does vanish at $q^2=0$.

>From the above expressions, we can see that the effect of operators
contributing to $\epsilon_1$ and $\epsilon_b$ is enhanced by a factor
$m^2_{\text{top}}/M_W^2$. This is in agreement with the results of
Ref.\ \cite{gouna} that used anomalous top interactions that transform
linearly under the action of $G$. Moreover, the right--handed charged
currents do not contribute to any of the observables and therefore
cannot be constrained by the LEPI data. Notice that the $\epsilon$
parameters depend on different combinations of the anomalous
couplings, providing a way to disentangle them in case of a clear sign
of new physics.

Our next step towards obtaining the bounds on the anomalous quartic
vertices is to determine the SM contribution to $\epsilon's$.  The
gauge-boson contribution to these parameters is infinite as a
consequence of the absence of the elementary Higgs. On the other hand,
one must also include the tree level contributions from the purely
gauge chiral Lagrangian \cite{appelquist}, which absorb
these infinities through the renormalization of the corresponding
constants.  If the renormalization condition is imposed at a scale
$\Lambda$, we are left with the contribution due to the running of the
couplings from the scale $\Lambda$ to the scale $\mu$. Therefore, the
SM contribution without the Higgs boson will be the same as that of
the SM with an elementary Higgs, with the substitution $ \ln
(M_H)\rightarrow \ln(\Lambda)$ \cite{dittmaier}.

We show in Table \ref{table} the 99\% CL constraints on the anomalous
top--quark interactions assuming that $\Lambda = 1$ TeV for $160$ GeV $\leq
m_{\text{top}}\leq 190$ GeV, provided that only one operator is
considered different from zero at each time.  In order to obtain these
bounds we constructed the $\chi^2$ function with the four epsilons
including the corresponding correlations.  The values shown in this
table verify the condition $\chi^2(m_{\text{top}},c_i,c_{j\neq i}=0)
\leq \chi^2_{min}(m_{\text{top}}, c_i,c_{j\neq i}=0)+ 6.7$ where $c_i$ is
the coefficient allowed to be different from zero at each time.  Our
results show that most operators get bounds close to the theoretical
expectation for their anomalous couplings, {\it i.e.} the bounds are
of order 1. However, there is an uncertainty in the derived bounds
associated with the choice for the $\mu$ scale, being the bounds in
Table \ref{table} derived for $\mu=m_{\text{top}}$.  Allowing $\mu=2
m_{\text{top}} \; (m_{\text{top}}/2)$ we get limits which are 10--20\%
weaker (stronger) than the ones given in this table.

\begin{figure}
\begin{center}
\mbox{\epsfig{file=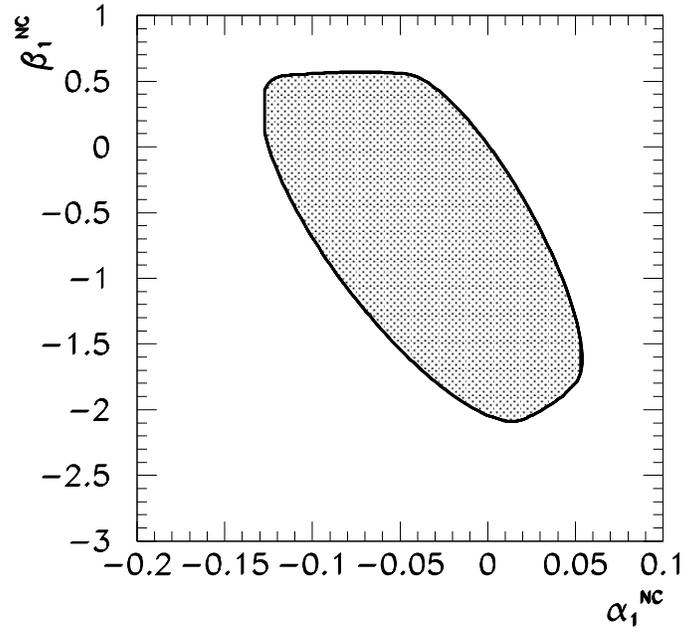,width=0.6\textwidth,height=0.4\textheight}}
\end{center} 
\caption{99\% CL allowed region for the parameters $\alpha_1^{NC}$ and
  $\beta_1^{NC}$ for for $\Lambda=1$ TeV and $160$ GeV $\leq
  m_{\text{{top}}}\leq 190$ GeV and $\mu=m_{\text{{top}}}$.}
\label{fig:1} 
\end{figure}

As a matter of fact, because of the large number of anomalous
couplings involved one can only obtain constraints on the different
combinations that contribute to each of the epsilon parameters. For
instance, we get for $m_{\text top}=170$ GeV that the regions allowed
at 99\% CL are
\begin{equation}
\begin{array}{c}
-0.105\leq \alpha_1^{NC}-\alpha_2^{NC}+0.38\gamma^{NC}-0.25
\gamma_{L}^{CC}\leq0.053 \\
-0.72\leq \beta_{1L}^{CC}-0.12\beta^1_{NC}-0.5\beta_2^{NC}
+0.16\gamma^{NC}-0.083 \gamma_{L}^{CC}\leq 0.53 \\
-2.2\leq \beta_1^{NC}+0.58\beta_2^{NC}-0.19\gamma^{NC}\leq 0.47 \\
-5.3\leq \alpha_{1L}^{CC}-4.6\beta_{1L}^{CC}+6\beta_{2L}^{CC}+7.9\gamma^{NC}
-0.41\gamma_L^{CC}+5.5\alpha_{2L}^{CC}\leq 8.2
\end{array}
\end{equation}
Moreover, there is also a large correlation between those parameters
which contribute to $\epsilon_1$ and $\epsilon_3$. For the sake of
illustration, we show in Fig.\ \ref{fig:1} the allowed region at 99\%
CL for the parameters $\alpha_1^{NC}$ and $\beta_1^{NC}$. 

Summarizing, we have analyzed the effects of possible anomalous
couplings between the top quark and the gauge bosons that appear in a
scenario where there is no particle associated to the
symmetry--breaking sector in the low--energy spectrum.  Using a chiral
Lagrangian formalism, we have constructed the most general
dimension--five CP invariant Lagrangian for the interactions between
the Goldstone bosons and the top and bottom quarks, which contains
seventeen unknown parameters. We then draw the limits on those
couplings arising from precision measurements at the $Z$ pole. Our
results show that right--handed charged currents do not contribute to
the LEPI observables and therefore cannot be constrained. We found
that left--handed charged-- and neutral--current contributions to
$\epsilon_1$ and $\epsilon_b$ are enhanced by a factor $m^2_{\text
{top}}/M_W^2$ .  Our limits on these operator are close to the
theoretically expected order of magnitude for these couplings.


\acknowledgments
We thank G. Altarelli and F. Caravagios for providing us with the 
new values of the epsilon parameters and their correlation matrix.
M.\ C.\ Gonzalez--Garcia is grateful to the Instituto de F\'{\i}sica
Te\'orica from Universidade Estadual Paulista for its kind hospitality.
We would like to thank A.\ Brunstein for discussions in the early stage
of this project. This work was supported by FAPESP (Brazil), CNPq (Brazil), 
DGICYT (Spain) under grant PB95--1077, and by CICYT (Spain) under grant
AEN96--1718.



\begin{table}
\begin{displaymath}
\protect
\begin{array}{||c||c||c||c||}
\hline\hline
\alpha_1^{NC} & -0.15\, , \,  0.013  &  \alpha_{1L}^{CC} & -1.6\, , \,  5.4 \\
 \alpha_2^{NC} & -0.013\, , \,  0.15 &\alpha_{2L}^{CC} & -0.40\, , \,  1.3\\
\beta_1^{NC} & -2.2\, , \,  0.15    & \beta_{L1}^{CC} & -0.65\, , \,  0.29\\
 \beta_2^{NC}  & -1.3\, , \,  0.44  & \beta_{L2}^{CC} & -0.26\, , \,  0.88  \\
 \gamma^{NC} &  -0.017\, , \,  0.20  &\gamma_{L}^{CC} & -0.56\, , \,  0.052  \\
\hline\hline
\end{array}
\end{displaymath}
\caption{99\% CL limits on the anomalous top couplings for $\Lambda=1$ TeV,
$160$ GeV $\leq m_{\text{{top}}}\leq 190$ GeV and $\mu=m_{\text{{top}}}$.}
\label{table}
\end{table}
\end{document}